# A General Implicit Splitting for Stabilizing Numerical Simulations of Langevin Equations

W. P. Petersen

Swiss Center for Scientific Computing, ETH, Zürich

**Abstract**

In this paper is described a general $2^{nd}$ order accurate (weak sense) procedure for stablizing Monte-Carlo simulations of Itô stochastic differential equations. The splitting procedure includes explicit Runge-Kutta methods [1], semi-implicit methods [2][3], and trapezoidal rule [1][4]. We prove the semi-implicit method of Öttinger [3] and note that it may be generalized for arbitrary splittings.

## 1 Introduction

We are interested in numerical procedures for simulating long time integrations of large dimensional Itô stochastic processes. The so-called Langevin dynamics approach of estimating physical quantities $<f>$ uses a vector of stochastic processes $\{x^\alpha(t), \alpha = 1, \ldots, n\}$ which vary with simulated time $t$ and converge to a stationary state (e.g. Parisi and Wu [5], Klauder [6]). Estimates for $<f>$ are long-time averages (large $T$)

$$<f> \approx \frac{1}{T} \int_0^T dt f(x(t)), \tag{1}$$

where process $x(t)$ satisfies a Langevin equation (Itô stochastic differential equation) of *multiplicative noise* form

$$dx^\alpha = b^\alpha(x)dt + \sigma^{\alpha\beta}(x)\, d\omega^\beta(t), \tag{2}$$

driven by an $n$-dimensional vector of independent Brownian motions $\omega$. Our notation in (2) uses the *summation convention* wherein repeated indices, in this case the $\beta = 1, \ldots, n$, are always summed over. Additionally, the following shorthand notation for partial derivatives will be found convenient: $\partial_\alpha g \equiv \partial g / \partial x^\alpha$ (for some function $g(x)$). In order that the long-time accuracy of (1) be of order of the time step size $h$, we desire that the single time step behavior of simulations of $x(t)$ be of $O(h^2)$.

Increments of the $n$ independent Brownian motions $\omega^\beta(t)$ satisfy the relations ($\alpha, \beta = 1, \ldots, n$, where $n$ is the size of vectors $x$ and $\omega$, and $\delta(t)$ is the Dirac delta function):

$$\begin{aligned}
\omega^\alpha(0) &= 0 \qquad (3)\\
<d\omega^\alpha(t)> &= 0\\
<d\omega^\alpha(t_1)\,d\omega^\beta(t_2)> &= \delta^{\alpha\beta}\,\delta(t_1-t_2)\,dt_1 dt_2\\
<d\omega^\alpha(t_1)\,d\omega^\beta(t_2)\,d\omega^\gamma(t_3)\,d\omega^\delta(t_4)> &= \delta^{\alpha\beta}\delta^{\gamma\delta}\,\delta(t_1-t_2)\,\delta(t_3-t_4)\,dt_1 dt_2 dt_3 dt_4\\
&+ \delta^{\alpha\gamma}\delta^{\beta\delta}\,\delta(t_1-t_3)\,\delta(t_2-t_4)\,dt_1 dt_2 dt_3 dt_4\\
&+ \delta^{\alpha\delta}\delta^{\beta\gamma}\,\delta(t_1-t_4)\,\delta(t_2-t_3)\,dt_1 dt_2 dt_3 dt_4
\end{aligned}$$

where the last relation follows from the third since the processes are Gaussian. These will suffice for all our needs to $O(h^2)$.

The stochastic integral over $\sigma(x)$ depending on $x$ must be handled carefully since the Brownian motion (Wiener process) $\omega(t)$ is not of bounded variation. In this paper we choose the Itô definition wherein (2) is shorthand for

$$x^\alpha(t) = x^\alpha(0) + \int_0^t b^\alpha(x(s))ds + \int_0^t \sigma^{\alpha\beta}(x(s))d\omega^\beta(s), \qquad (4)$$

and the stochastic integral is belated, an Itô martingale [12]. In simple terms, the belated integral may be considered a Riemann sum in which the value taken for integrand $\sigma^{\alpha\beta}(x(s))$ in each $t$-interval ($t_j \leq s < t_{j+1}$) is for argument $x(t_j)$ at the beginning of the interval [6]. Other definitions, e.g. Stratonovich, are equivalent by transformations of the drift $\int b\,ds$ under sufficient smoothness conditions on $\sigma$ (e.g. [12]).

## 2 Numerical Approximations

We begin by stating our algorithm and put off discussions of variants until later. For (2), writing $b(x)$ split into two parts

$$b^\alpha(x) = A^\alpha(x) + B^\alpha(x),$$

the following is a second order accurate (weak sense) simulation method. Vector $x_h$ is the process value at the end of time step $h$, and $x_0$ is the process value at the beginning of the step.

$$\begin{aligned}
x_h^\alpha = x_0^\alpha &+ \frac{h}{2}(A^\alpha(x_h) + B^\alpha(x_0 + \sigma_0\xi_1 + (A_0+B_0)h) + A^\alpha(x_0) + B^\alpha(x_0))\\
&+ \frac{1}{2}\{\sigma^{\alpha\beta}(x_0 + \sqrt{\frac{1}{2}}\sigma_0\xi_0 + \frac{h}{2}(A_0+B_0)) \qquad (5)\\
&\quad + \sigma^{\alpha\beta}(x_0 - \sqrt{\frac{1}{2}}\sigma_0\xi_0 + \frac{h}{2}(A_0+B_0))\}\xi_1^\beta\\
&+ (\partial_\beta\sigma_0^{\alpha\delta})\,\sigma_0^{\beta\epsilon}\,\Xi^{\epsilon\delta}.
\end{aligned}$$

In this formula (5), numbers $\xi_0$ and $\xi_1$ are vectors of independent, identically distributed (iid), Gaussian (to $O(h^2)$) random variables with zero mean and variance $h$:

$$
\begin{aligned}
<\xi_0^\alpha> &= <\xi_1^\alpha> = 0, \\
<\xi_0^\alpha \xi_1^\beta> &= 0, \\
<\xi_0^\alpha \xi_0^\beta> &= <\xi_1^\alpha \xi_1^\beta> = \delta^{\alpha\beta} h, \\
<\xi_i^\alpha \xi_i^\beta \xi_i^\gamma \xi_i^\delta> &= \left(\delta^{\alpha\beta}\delta^{\gamma\delta} + \delta^{\alpha\gamma}\delta^{\beta\delta} + \delta^{\alpha\delta}\delta^{\beta\delta}\right) h^2, \quad i = 0, 1.
\end{aligned}
\tag{6}
$$

Again, the last relation follows from from the third since the $\xi_0$ and $\xi_1$ are Gaussian. The $\xi_i$ are well modeled by $\xi_i^\alpha = \sqrt{h} z_i^\alpha$, where $z_i^\alpha$ are $2 \cdot n$ ($\alpha = 1, \ldots, n$ and $i = 0, 1$) normal random variates, each with zero mean and unit variance [1],[8],[9]. Array $\Xi^{\epsilon\delta}$ contains models for the stochastic integrals (see Talay [11] for example, or Kloeden and Platen [6]):

$$
\Xi^{\epsilon\delta} \approx \int_t^{t+h} \omega^\epsilon(s) d\omega^\delta(s) = O(h).
$$

These will be discussed in more detail in our proof of (5). Additionally, subscript 0 on vector $x_0$ refers to the value of the process at the beginning of the time step $t \to t + h$, and on the drift/diffusion coefficients indicate $A_0 = A(x_0)$, $B_0 = B(x_0)$, etc.

## 2.1 Taylor series and stochastic integrals

In this section we set out some known results (e.g. Milstein book [9]) for the purposes of explaining notation and to make the arguments coherent. The idea of weak numerical approximations to (2) involves constructing a sequence of discrete $x_{l \cdot h}$ $l = 0, \ldots$ vectors, one for each time step. Then, if $x_{l \cdot h} = x(t)$, for an arbitrary smooth function $f$ ($C^4$ in each $x^\alpha$), we want the functionals

$$
\underbrace{< f(x(t+h)) >}_{\text{computed using (3)}} = \underbrace{< f(x_{(l+1) \cdot h}) >}_{\text{computed using (6)}}
$$

to agree to some desired order in $h$. More precisely, if at time $t$, process $x(t)$ has value $x_0$ and $f(x(t))$ has value $f_0$, we require that

$$\begin{aligned}
< f(x(t+h)) > &= f_0 + \frac{\partial f_0}{\partial x^\alpha} < \Delta x^\alpha > \\
&\quad + \frac{1}{2}\frac{\partial^2 f_0}{\partial x^\alpha \partial x^\beta} < \Delta x^\alpha \Delta x^\beta > \\
&\quad + \frac{1}{3!}\frac{\partial^3 f_0}{\partial x^\alpha \partial x^\beta \partial x^\gamma} < \Delta x^\alpha \Delta x^\beta \Delta x^\gamma > \\
&\quad + \frac{1}{4!}\frac{\partial^4 f_0}{\partial x^\alpha \partial x^\beta \partial x^\gamma \partial x^\delta} < \Delta x^\alpha \Delta x^\beta \Delta x^\gamma \Delta x^\delta > \\
&= < f(x_{t+h}) > + o(h^2).
\end{aligned} \qquad (7)$$

We compute the moments $< \Delta x^\alpha \cdots \Delta x^\delta >$ from expectation values of products of some stochastic integrals found when one step of (4) is expanded in a Taylor series. These moments must agree to the desired order in $h$ ($O(h^2)$ in our case) with moments computed using the simulation (5). This idea, apparently due to Wagner and Platen [15], is well described in Milstein [9] or Kloeden and Platen [8]. That is, we write

$$\Delta x^\alpha = \underbrace{\int_t^{t+h} b^\alpha(x(s))ds}_{\Delta \mathcal{A}^\alpha(h)} + \underbrace{\int_t^{t+h} \sigma^{\alpha\beta}(x(s))d\omega^\beta(s)}_{\Delta \mathcal{M}^\alpha(h)}, \qquad (8)$$

where $\Delta\mathcal{M}^\alpha(h) = O(h^{\frac{1}{2}}) + O(h) + \cdots$, and $\Delta\mathcal{A}^\alpha(h) = O(h) + O(h^{\frac{3}{2}}) + \cdots$, and $x(s)$ appearing on the right hand side of (8) may be repeatedly substituted (in Picard fashion) by the right hand side of the integral formula (4). The result is a stochastic Taylor series (e.g. see [8], Chapter 5). If we can construct discrete models for the resultant stochastic integrals to $O(h^2)$, these models may be inserted into the stochastic Taylor series for $\Delta x^\alpha$ to obtain a model Taylor series whose increments satisfy (7).

## 2.2 Stochastic Taylor series

We first derive the stochastic Taylor series to $O(h^2)$, then find some models for the required stochastic integrals. These results may be found elsewhere in more general formulations (e.g. [8]). It is important to note that $x^\alpha(t)$ is a continuous Markov process and that our simulation is a discrete one. A method for computing any one step enables us to compute any other step. Hence, without loss of generality, the integrals from $t \to t+h$ my be abbreviated to $0 \to h$. For example, consider the following step of the stochastic Taylor series:

$$\int_t^{t+h} \sigma(x(s))d\omega(s) = \int_t^{t+h} \sigma(x(t) + \Delta x) \, d\omega(s)$$

$$= \sigma_0 \int_t^{t+h} d\omega(s) + (\partial_x \sigma_0)\sigma_0 \int_t^{t+h} \left\{ \int_t^s d\omega(u) \right\} d\omega(s) + ...$$

$$= \sigma_0 \Delta\omega_h + (\partial_x \sigma_0)\sigma_0 \int_t^{t+h} \Delta\omega_s d\omega(s) + ...$$

Now $\Delta\omega_s = \int_t^{t+s} d\omega(u)$ is a finite increment of Brownian motion $\omega$ and has initial value $\Delta\omega_0 = 0$. Furthermore, the infinitesimal increments (in $s$) of $\Delta\omega_s$ are $d\omega(s)$, whose properties are (3). Thus, $\Delta\omega_s$ may be treated as a Brownian motion in the finite interval $t \leq t + s \leq t + h$, i.e. $0 \leq s \leq h$, of interest.

The two terms of (8) are

$$\Delta \mathcal{A}^\alpha(h) = b_0^\alpha h + (\partial_\beta b_0^\alpha) \, \sigma_0^{\beta\gamma} \underbrace{\int_0^h ds\, \omega^\gamma(s)}_{J^\gamma} \tag{9}$$

$$+ (\partial_\beta b_0^\alpha)(\partial_\gamma \sigma_0^{\beta\epsilon})\sigma_0^{\gamma\iota} \underbrace{\int_0^h ds \int_0^s \omega^\iota(u)\, d\omega^\epsilon(u)}_{*}$$

$$+ \frac{1}{2}(\partial_\beta b_0^\alpha)\, b_0^\beta\, h^2 + \frac{1}{2}(\partial_\beta \partial_\gamma b_0^\alpha)\sigma_0^{\beta\epsilon}\sigma_0^{\gamma\iota} \underbrace{\int_0^h ds\, \omega^\epsilon(s)\omega^\iota(s)}_{K^{\epsilon\iota}} + o(h^2)$$

and

$$\Delta \mathcal{M}^\alpha(h) = \sigma_0^{\alpha\beta} \underbrace{\int_0^h d\omega^\beta(h)}_{I^\alpha} + (\partial_\beta \sigma_0^{\alpha\gamma})\sigma_0^{\beta\epsilon} \underbrace{\int_0^h \omega^\epsilon(s)\, d\omega^\gamma(s)}_{I^{\epsilon\gamma}} \tag{10}$$

$$+ (\partial_\beta \sigma_0^{\alpha\gamma})(\partial_\iota \sigma_0^{\beta\epsilon})\sigma_0^{\iota\eta} \underbrace{\int_{s=0}^h \left(\int_{u=0}^s \omega^\eta(u)d\omega^\iota(u)\right) d\omega^\gamma(s)}_{I^{\eta\iota\gamma}}$$

$$+ (\partial_\beta \sigma_0^{\alpha\gamma})\, b_0^\beta \underbrace{\int_0^h s\, d\omega^\gamma(s)}_{K^\gamma}$$

$$+ \frac{1}{2}(\partial_\beta \partial_\kappa \sigma_0^{\alpha\gamma})\sigma_0^{\beta\gamma}\sigma_0^{\kappa\epsilon} \underbrace{\int_0^h \omega^\sigma \omega^\epsilon\, d\omega^\gamma}_{L^{\sigma\epsilon\gamma}} + \underbrace{O(h^2)}_{*}$$

In these expressions, those terms underbrace marked with $*$ are $O(h^2)$ with vanishing expectation values and may be ignored. That is, in (7) those $O(h^2)$ terms whose expectation values are zero cannot contribute to any of the $<\Delta x \cdots \Delta x>$ moments to $O(h^2)$. In particular, **all** $O(h^2)$ terms of the martingale $\Delta \mathcal{M}^\alpha(h)$ may be ignored. The remaining labeled stochastic integrals are modeled as follows.

## 2.3 Models for Stochastic Integrals

In increasing order of powers of $h$ (the step size), the needed stochastic integrals and their models are below. Following the enumeration, we prove the models only for $J^\gamma$ and $L^{\sigma\epsilon\gamma}$. Model $\Xi^{\epsilon\gamma}$ is shown in [8] and [11] and like the remainder is a straightforward application of the correlations (3) and showing that products of the models have the same expectation values as their corresponding stochastic integrals to $O(h^2)$ accuracy. To verify the model for $\Xi^{\epsilon\gamma}$, only products $\xi_1^\alpha$ and $\xi_1^\beta$ times $\Xi^{\epsilon\gamma}$, and $\Xi^{\alpha\beta}$ times $\Xi^{\epsilon\gamma}$ need be considered. No other terms of $\Delta x$ in (8) can form any contribution to (7) to $O(h^2)$.

$O(h^{\frac{1}{2}})$:

$$I^\alpha = \Delta\omega_h^\alpha = \int_0^h d\omega^\alpha(s) \approx \xi_1^\alpha = \sqrt{h} z_1^\alpha$$

$O(h^1)$:

$$\begin{aligned}
I^{\epsilon\gamma} = \int_0^h \omega^\epsilon(s)\, d\omega^\gamma(s) &\approx \Xi^{\epsilon\gamma} \\
&= \frac{h}{2}\left(z_1^\epsilon z_1^\gamma - \tilde{z}^{\epsilon\gamma}\right) \quad \epsilon > \gamma \\
&= \frac{h}{2}\left(z_1^\epsilon z_1^\gamma + \tilde{z}^{\gamma\epsilon}\right) \quad \epsilon < \gamma \\
&= \frac{h}{2}\left((z_1^\epsilon)^2 - 1\right) \quad \epsilon = \gamma
\end{aligned}$$

Here, variables $z_1^\epsilon, z_1^\gamma$ are the same zero mean, unit variance Gaussians that appear in the model for $\xi_1^\epsilon$ (respectively $\xi_1^\gamma$). Array $\tilde{z}^{\epsilon\gamma}$ is defined only for $\epsilon > \gamma$ and consists of $n \cdot (n-1)/2$ independent, zero mean, unit variance normal (to $O((\tilde{z})^4)$) random variables. These are independent of the $z_1$, $z_0$, and each other.

$O(h^{\frac{3}{2}})$:

$$J^\gamma = \int_0^h ds\,\omega^\gamma(s) \approx \frac{1}{2}h\,\xi_1^\gamma$$

$$K^\gamma = \int_0^h s\,d\omega^\gamma(s) \approx \frac{1}{2}h\,\xi_1^\gamma$$

$$I^{\eta\iota\gamma} = \int_{s=0}^h \left(\int_{u=0}^s \omega^\eta(u)d\omega^\iota(u)\right) d\omega^\gamma(s) \approx 0$$

$$L^{\sigma\epsilon\gamma} = \int_0^h \omega^\sigma \omega^\epsilon\,d\omega^\gamma \approx \frac{1}{2}h\,\delta^{\sigma\epsilon}\,\xi_1^\gamma,$$

$$\text{or} \approx \frac{1}{2}\xi_0^\sigma \xi_0^\epsilon \xi_1^\gamma$$

In $L^{\sigma\epsilon\gamma}$, the variables $\xi_0^\sigma = \sqrt{h}\,z_0^\sigma$ contain independent zero mean, unit variance Gaussians $z_0^\sigma$: independent of the $z_1^\alpha$ appearing in $\xi_1^\alpha$ and the $\tilde{z}^{\epsilon\gamma}$ appearing in $\Xi^{\epsilon\gamma}$.

$O(h^2)$:

$$K^{\epsilon\iota} = \int_0^h ds\,\omega^\epsilon(s)\omega^\iota(s) \approx \frac{h}{2}\xi_1^\epsilon\xi_1^\iota,$$

$$\text{or} \approx \frac{h^2}{2}\delta^{\epsilon\iota},$$

$$\text{or} \approx \frac{h}{2}\xi_0^\epsilon\xi_0^\iota$$

All three models for $K^{\epsilon\iota}$ satisfy the calculus to $O(h^2)$.

**Proof of model for $J^\gamma$:**

Since $J^\gamma \approx \frac{h}{2}\xi_1^\gamma$ is $O(h^{\frac{3}{2}})$ with vanishing expectation, we observe that only products of $J^\gamma$ with terms of $O(h^{\frac{1}{2}})$ can contribute to $O(h^2)$ in (7). Respectively then,

- both $J^\gamma$ and the model $\frac{h}{2}\xi_1^\gamma$ have vanishing expectation, so the calculus is satisfied to $O(h^{\frac{3}{2}})$.
- The expectation of product $J^\gamma$ times $I^\alpha$ is

$$<I^\alpha J^\gamma> = <(\int_{u=0}^h d\omega^\alpha(u))(\int_{s=0}^h ds(\int_{v=0}^s d\omega^\gamma(v)))>$$

$$= \int_{u=0}^h du \int_{s=0}^h ds \int_{v=0}^s dv\,\delta^{\alpha\gamma}\,\delta(u-v)$$

$$= \frac{h^2}{2}\delta^{\alpha\gamma}$$

while the product of the models $\xi_1^\alpha$ and $\frac{h}{2}\xi_1^\gamma$ has expectation (in the space of $z$'s):

$$< \xi_1^\alpha \frac{h}{2} \xi_1^\gamma > = \frac{h^2}{2} < z_1^\alpha z_1^\gamma > = \frac{h^2}{2} \delta^{\alpha\gamma}$$

where (6) has been used.

**Proof of model for $L^{\sigma\epsilon\gamma}$:**

Since $L^{\sigma\epsilon\gamma}$ and the its proposed models are $O(h^{\frac{3}{2}})$, we observe that

- both $L^{\sigma\epsilon\gamma}$ and the models have vanishing expectations. These follow from the facts that $L^{\sigma\epsilon\gamma}$ is a martingale, and that the models are odd in in the Gaussian variables $\xi_1$, respectively. Thus, to $O(h^{\frac{3}{2}})$, the calculus is satisfied.

- Inserting

$$\omega^\sigma(t) = \int_{u=0}^{t} d\omega^\sigma(u), \quad \text{and}$$

$$\omega^\epsilon(t) = \int_{v=0}^{t} d\omega^\epsilon(v)$$

into the expectation of product $L^{\sigma\epsilon\gamma}$ times $I^\alpha$ we get

$$< I^\alpha L^{\sigma\epsilon\gamma} > = < (\int_{s=0}^{h} d\omega^\alpha(s))(\int_{t=0}^{h} \omega^\sigma(t)\omega^\epsilon(t)\,d\omega^\gamma(t)) >$$

$$= \int_{s=0}^{h} ds \int_{t=0}^{h} dt \int_{v=0}^{t} dv \int_{u=0}^{t} du$$
$$\{\delta^{\alpha\sigma}\delta^{\epsilon\gamma}\delta(s-u)\delta(t-v)$$
$$+ \delta^{\alpha\epsilon}\delta^{\sigma\gamma}\delta(s-v)\delta(t-u)$$
$$+ \delta^{\alpha\gamma}\delta^{\sigma\epsilon}\delta(s-t)\delta(u-v)\}$$

$$= \frac{h^2}{2}\delta^{\alpha\gamma}\delta^{\sigma\epsilon}$$

In the four-fold integral: since $v < t$, the first term vanishes, and likewise $u < t$ eliminates the second, so only the last term survives. The product of the model $\xi_1^\alpha$ and the first model $\frac{h}{2}\delta^{\sigma\epsilon}\xi_1^\gamma$ has expectation (in the space of $z$'s):

$$< \xi_1^\alpha \frac{h}{2}\delta^{\sigma\epsilon}\xi_1^\gamma > = \frac{h^2}{2}\delta^{\sigma\epsilon} < z_1^\alpha z_1^\gamma > = \frac{h^2}{2}\delta^{\alpha\gamma}\delta^{\sigma\epsilon}.$$

Likewise, for the second proposed model,

$$< \xi_1^\alpha \frac{1}{2}\xi_0^\sigma \xi_0^\epsilon \xi_1^\gamma > = \frac{h^2}{2} < z_0^\sigma z_0^\epsilon >< z_1^\alpha z_1^\gamma > = \frac{h^2}{2}\delta^{\alpha\gamma}\delta^{\sigma\epsilon}$$

where (6) has been used to compute the expectations in both variants. Thus, the models for $L^{\sigma\epsilon\gamma}$ satisfy the calculus to $O(h^2)$.

## 2.4 Simplified Taylor series

Substituting the models for stochastic integrals of section 2.3 into the Taylor series of section 2.2, we get a **model** Taylor series

$$
\begin{aligned}
x_h^\alpha &= x_0^\alpha + & (11) \\
&+ b_0^\alpha h + (\partial_\beta b_0^\alpha) \sigma_0^{\beta\gamma} \frac{h}{2} \xi_1^\gamma & \text{drift } \Delta\mathcal{A}^\alpha(h) \\
&+ \frac{1}{2} \left( (\partial_\beta b_0^\alpha) b_0^\beta h + \frac{1}{2} (\partial_\beta \partial_\gamma b_0^\alpha) \sigma_0^{\beta\epsilon} \sigma_0^{\gamma\iota} \xi_1^\epsilon \xi_1^\iota \right) h \\
&+ \sigma_0^{\alpha\beta} \xi_1^\beta + (\partial_\beta \sigma_0^{\alpha\gamma}) \sigma_0^{\beta\epsilon} \Xi^{\epsilon\gamma} & \text{diffusion } \Delta\mathcal{M}^\alpha(h) \\
&+ \frac{1}{2} \left( (\partial_\beta \sigma_0^{\alpha\gamma}) b_0^\beta h + \frac{1}{2} (\partial_\beta \partial_\kappa \sigma_0^{\alpha\gamma}) \sigma_0^{\beta\sigma} \sigma_0^{\kappa\epsilon} \xi_0^\sigma \xi_0^\epsilon \right) \xi_1^\gamma
\end{aligned}
$$

## 3 Proof of splitting formula

Using the model Taylor series (11), we now verify the splitting formula (5). This is straightforward, writing for brevity (5) as

$$\Delta x^\alpha = \frac{h}{2} \{ A^\alpha(x_h) + B^\alpha(x_{euler}) + A^\alpha(x_0) + B^\alpha(x_0) \} + \Delta\mathcal{M}^\alpha(h) \quad (12)$$

where $x_{euler} = x_0 + hb_0 + \sigma_0 \xi_1$ is the Euler estimate, and

$$
\begin{aligned}
A^\alpha(x_h) &= A^\alpha(x_0 + \Delta x) \\
&= A_0^\alpha + (\partial_\beta A_0^\alpha) \Delta x^\beta + \frac{1}{2} (\partial_\beta \partial_\gamma A_0^\alpha) \Delta x^\beta \Delta x^\gamma + o(h)
\end{aligned}
$$

is all we need to $O(h^2)$. Expanding the second term on the right hand side one more time

$$
\begin{aligned}
(\partial_\beta A_0^\alpha) \Delta x^\beta &= (\partial_\beta A_0^\alpha) \left\{ h(A_0^\beta + B_0^\beta) + \Delta\mathcal{M}^\beta(h) \right\} + o(h) \\
&= (\partial_\beta A_0^\alpha) \left\{ h(A_0^\beta + B_0^\beta) + \frac{1}{2} (\sigma^{\beta\gamma}(x_+) + \sigma^{\beta\gamma}(x_-)) \xi_1^\gamma + (\partial_\delta \sigma_0^{\beta\epsilon}) \sigma_0^{\delta\iota} \Xi^{\iota\epsilon} \right\} \\
&\quad + o(h)
\end{aligned}
$$

The $O(h)$ term containing $\Xi^{\iota\epsilon}$ may be ignored since it has vanishing expectation. The whole expression containing $(\partial_\beta A_0^\alpha) \Delta x^\beta$ is already $O(h)$, so this term will be $O(h^2)$ overall. Subscripts on $x$, $x_+$ and $x_-$, are the arguments

$$x_+^\kappa = x_0^\kappa + \sqrt{\frac{1}{2}}\sigma_0^{\kappa\lambda}\xi_0^\lambda + \frac{h}{2}(A_0^\kappa + B_0^\kappa)$$

$$x_-^\kappa = x_0^\kappa - \sqrt{\frac{1}{2}}\sigma_0^{\kappa\lambda}\xi_0^\lambda + \frac{h}{2}(A_0^\kappa + B_0^\kappa).$$

Because

$$\sigma^{\beta\gamma}(x_+) + \sigma^{\beta\gamma}(x_-) = 2\sigma_0^{\beta\gamma} + O(h),$$

by way of the explicit construction of $x_+$ and $x_-$, we have

$$(\partial_\beta A_0^\alpha)\Delta x^\beta = (\partial_\beta A_0^\alpha)\left\{h(A_0^\beta + B_0^\beta) + \sigma_0^{\beta\gamma}\xi_1^\gamma\right\} + o(h).$$

Including this expansion in (12),

$$\Delta x^\alpha = \frac{h}{2}\left\{A_0^\alpha + (\partial_\beta A_0^\alpha)(A_0^\beta + B_0^\beta)h + (\partial_\beta A_0^\alpha)\sigma_0^{\beta\gamma}\xi_1^\gamma + (\partial_\beta\partial_\gamma A_0^\alpha)\sigma_0^{\beta\delta}\sigma_0^{\gamma\epsilon}\xi_1^\delta\xi_1^\epsilon + B^\alpha(x_{euler}) + A^\alpha(x_0) + B^\alpha(x_0)\right\} + \Delta\mathcal{M}^\alpha(h)$$

leaving only the $B(x_{euler})$ term left. This is

$$B^\alpha(x_{euler}) = B^\alpha(x_0 + b_0 h + \sigma_0 \xi_1)$$

where $b_0 = A_0 + B_0$. We now need to expand this to $O(h)$,

$$B^\alpha(x_{euler}) = B_0^\alpha + (\partial_\beta B_0^\alpha)(A_0^\beta + B_0^\beta)h + (\partial_\beta B_0^\alpha)\sigma_0^{\beta\gamma}\xi_1^\gamma + \frac{1}{2}(\partial_\beta\partial_\gamma B_0^\alpha)\sigma_0^{\beta\epsilon}\sigma_0^{\gamma\eta}\xi_1^\epsilon\xi_1^\eta + o(h)$$

Or, altogether,

$$\Delta x^\alpha = \frac{h}{2}\left\{A_0^\alpha + (\partial_\beta A_0^\alpha)(A_0^\beta + B_0^\beta)h + (\partial_\beta A_0^\alpha)\sigma_0^{\beta\gamma}\xi_1^\gamma \right.$$
$$+ (\partial_\beta\partial_\gamma A_0^\alpha)\sigma_0^{\beta\delta}\sigma_0^{\gamma\epsilon}\xi_1^\delta\xi_1^\epsilon +$$
$$B_0^\alpha + (\partial_\beta B_0^\alpha)(A_0^\beta + B_0^\beta)h + (\partial_\beta B_0^\alpha)\sigma_0^{\beta\gamma}\xi_1^\gamma$$
$$\left. + \frac{1}{2}(\partial_\beta\partial_\gamma B_0^\alpha)\sigma_0^{\beta\epsilon}\sigma_0^{\gamma\eta}\xi_1^\epsilon\xi_1^\eta + A^\alpha(x_0) + B^\alpha(x_0)\right\} +$$
$$\Delta\mathcal{M}^\alpha(h)$$

$$= (A_0^\beta + B_0^\beta)h + (\partial_\beta(A_0^\alpha + B_0^\alpha))(A_0^\beta + B_0^\beta)\frac{h^2}{2} +$$
$$(\partial_\beta(A_0^\alpha + B_0^\alpha))\sigma_0^{\beta\gamma}\frac{h}{2}\xi_1^\gamma +$$
$$(\partial_\beta\partial_\gamma(A_0^\alpha + B_0^\alpha))\sigma_0^{\beta\epsilon}\sigma_0^{\gamma\eta}\xi_1^\epsilon\xi_1^\eta +$$
$$\Delta\mathcal{M}^\alpha(h)$$

Again using $b = A + B$, we get (11), which proves the formula (5).

## 4 Explicit and semi-implicit variants, stability

The significance of the splitting formula (5) lies principally in the increased stability of simulations when $A \neq 0$. There are a plethora of references to implicit methods in ordinary differential equations (ODE's), most of which really only treat linear stability. For example, Bulirsch and Stoer [13] or Gear [14] discuss a basic analysis of the linear ODE, $d\vec{x}(t) = A\vec{x}dt$, where $A$ is a negative matrix with eigenvalues having large scale differences. Namely, when $|\lambda_{max}|/|\lambda_{min}|$ is egregiously large, these scale differences in the eigenvalues ($\lambda$) of $A$ make simulations difficult. A time step $h$ small enough to resolve short time scales is much too small to be practical for long time scale components. Increasing the time step leads to unstable simulations. Indeed, this basic linear analysis may be extended to the stochastic differential equation case. The obvious analog is a vector version of the Ornstein-Uhlenbeck process: $d\vec{x} = A\vec{x}dt + d\vec{\omega}(t)$. Several analysis of this situation exist (e.g. [1], [10]), and will be only briefly discussed here. Instead, we will discuss some simple non-linear problems, where the drift is monomial.

### 4.1 Trapezoidal rule and semi-implicit methods

The solution of (5) can be a complicated affair when the drift is highly non-linear. At every time step, this equation must be solved for solution $x_h$ which appears on both sides of the algorithm. However, in a common case

$$b^\alpha(x) = A^{\alpha\beta}x^\beta + g^\alpha(x),$$

matrix $A$ forms a linear part of $b$, and $g(x)$ is non-linear. An obvious choice of splitting makes (5) easy to solve:

$$A^\alpha(x) = A^{\alpha\beta}x^\beta \qquad \text{and} \qquad B^\alpha(x) = g^\alpha(x).$$

This solution is affected by moving the $\frac{h}{2}Ax_h$ term which then appears on the right hand side of (5) to the left, computing $(1 - \frac{h}{2}A)^{-1}$, and solving a linear system where everything on the right hand side is explicit. Variations similar to this have been known for a long time in ODE simulations (e.g. see Butcher [16]). For this **semi-implicit** splitting, the linear stability properties discussed in, say Gear[14], are preserved, but the resulting implicit equations are easy to solve. Other variants include the following.

The choice
$$A^\alpha(x) = b^\alpha(x) \qquad \text{and} \qquad B^\alpha(x) = 0,$$
gives an **implicit trapezoidal rule**, while the alternative

$$B^\alpha(x) = b^\alpha(x) \qquad \text{and} \qquad A^\alpha(x) = 0,$$

is an **explicit trapezoidal rule**, and is a $2^{nd}$ order Runge-Kutta method. Schurz [4] has shown that in many cases, in particular when the drift $b = b(t)$ is not autonomous, some degree of implicitness can be essential. His most glaring example is the case of Brownian bridge, $x = \omega(t) - \frac{t}{t_1}\omega(t_1)$, where $x(0) = 0$ and $x(t_1) = 0$, and $t_0 \leq t \leq t_1$. For that case, when $A = \alpha b(t)$ and $B = (1-\alpha)b(t)$, only non-zero $\alpha$ (in fact, any non-zero $\alpha$) gives correct results. Additionally, he showed that only implicit trapezoidal rule is asymptotically un-biased.

Frequently, for example in polymer physics, it is the non-linear part of $b$ that causes instabilities. Öttinger [3] has used a splitting wherein a linear part of $b(x) = Ax + g(x)$, $B(x) = Ax$, is chosen for the explicit part, and the non-linear part, $g(x)$, is taken for the implicit term $A(x) = g(x)$. In the example in the next section (4.2), a monomial drift is illustrated wherein the increased stability is demonstrated.

## 4.2 Simple stability analysis for implicit algorithms

In simulations of ordinary differential equations, a method is said to be stable if when applied to the linear equation $\dot{x} = Ax$, the discrete solution $x(k \cdot h) \to 0$ as $k \to \infty$ when matrix $A < 0$. For Langevin equations, if $\sigma(x = 0) \neq 0$, the solution doesn't degenerate, that is, $x$ doesn't vanish even though $b$ is contracting. The analog to $x \to 0$ for the Langevin case is convergence to a strictly stationary process (e.g. see Doob [17]). Thus, we consider an approximate analysis of the following scalar additive noise problem

$$dx = b(x)dt + dw(t),$$

where $b$ is contracting. That is, $b(x)$ is skew in the sense that $b(x) < 0$ when $x >> 0$ and $b(x) > 0$ when $x << 0$. As $t \to \infty$, the distribution function for $x$ satisfies the forward Kolmogorov equation and becomes time independent:

$$\partial_t p(x,t) = \partial_x(\frac{1}{2}\partial_x - b(x))p(x,t) \to 0.$$

Thus, the stationary distribution function in this limit is ($x \geq 0$)

$$p(x, t \to \infty) = N \; exp(\; 2 \int_0^x b(z)dz), \tag{13}$$

where $N$ is a normalization. Let $\overline{|x|^2}_\infty = <|x|^2>_{t\to\infty}$ be the asymptotic mean square. A necessary condition for mean square stability is then

$$< |x_0 + \Delta x|^2 > \leq |x_0|^2 \tag{14}$$

whenever $|x_0|^2 >> \overline{|x|^2}_\infty$. In the infinitesimal $h$ limit, this condition for $b$ to be contracting when $|x_0|$ is large is

$$2 \,\mathbf{Re}\,(\overline{x}_0 b(x_0)) + 1 \leq 0.$$

For the discrete simulation, however, (14) is step size and process size dependent, just as in the ODE case. To illustrate the situation, we compare **explicit** and **implicit** trapezoidal rule algorithms. These are, from (5), respectively

$$x_h = x_0 + \frac{h}{2}(b(x_{euler}) + b(x_0)) + \xi, \tag{15}$$

where $x_{euler}$ is the Euler estimate (see Section 3), and

$$x_h = x_0 + \frac{h}{2}(b(x_h) + b(x_0)) + \xi. \tag{16}$$

The necessary condition (14) applied to the two methods yields an approximate, and as we hopefully demonstrate, qualitatively correct, analysis obtained by expanding (15) and (16) to $O(h^2)$. This inequality (14) for the explicit form (15) is

$$\begin{aligned}
< |x_h|^2 > - |x_0|^2 &= |x_0 + hb_0 + \frac{h^2}{2}b'_0 b_0|^2 \\
&\quad + \frac{h^2}{2}((x_0 + hb_0 + \frac{h^2}{2}b'_0 b_0)b''_0) \\
&\quad + \frac{3h^4}{16}|b''_0|^2 + |1 + \frac{h}{2}b'_0|^2 h - |x_0|^2 \\
&\leq 0.
\end{aligned} \tag{17}$$

We can get a semi-implicit approximation for the implicit trapezoidal rule as follows. First, we move all the $x_h$ dependent terms to the left hand side

$$x_h - \frac{h}{2}b(x_h) = x_0 + \frac{h}{2}b(x_0) + \xi,$$

to be expanded in a Taylor series in $\Delta x = x_h - x_0$. We get

$$\Delta x - \frac{h}{2}b'_0 \Delta x - \frac{h}{4}b''_0(\Delta x)^2 = hb_0 + \xi + o(h^2).$$

Now notice that $(\Delta x)^2 = (\xi)^2 + o(h)$, whence

$$x_h = x_0 + (1 - \frac{h}{2}b'_0)^{-1}\left\{hb_0 + \frac{h}{4}b''_0\xi^2 + \xi\right\}. \tag{18}$$

Squaring this expression and taking expectations produces a semi-implicit approximation for the inequality (14):

$$<|x_h|^2> - |x_0|^2 = 2(1 - \frac{h}{2}b'_0)^{-1}(hb_0 + \frac{h^2}{4}b''_0)x_0 + \tag{19}$$
$$|(1 - \frac{h}{2}b'_0)^{-2}(hb_0 + \frac{h^2}{4}b''_0)^2| + |1 - \frac{h}{2}b'_0|^{-2}h$$
$$\leq 0.$$

When a local linearization of the drift $b(x)$ is permissible, in particular the Ornstein-Uhlenbeck process $b(x) = b'_0 x$, these conditions reduce to

$$|1 + hb'_0 + \frac{h^2}{2}(b'_0)^2|^2|x_0|^2 + |1 + \frac{h}{2}b'_0|^2 h \leq |x_0|^2,$$

or

$$|1 + hb'_0 + \frac{h^2}{2}(b'_0)^2|^2 < 1. \tag{20}$$

And, for the implicit case:

$$\left|\frac{1 + \frac{h}{2}b'_0}{1 - \frac{h}{2}b'_0}\right|^2 |x_0|^2 + (1 - \frac{h}{2}b'_0)^{-2}h \leq |x_0|^2,$$

or

$$\left|\frac{1 + \frac{h}{2}b'_0}{1 - \frac{h}{2}b'_0}\right|^2 < 1. \tag{21}$$

We notice that although $b'_0 < 0$ ($b$ is contracting), whenever $|hb'_0|$ is large enough the first inequality (20) fails. However, for $b'_0 < 0$ the second (21) is satisfied for large step-sizes. Experiments described in [1] show these conclusions in more detail.

For non-linear problems where no Lipshitz bound on $b(x)$ exists, and therefore a local linearization says nothing about larger $x_0$ behavior, the analysis isn't quite so simple but seems to give the same conclusion. Namely, the implicit rule (16) is significantly more stable than the explicit one (15). To put a finer point on this we looked at some monomial drift problems

$$dx = -x|x|^{m-1}dt + dw(t), \tag{22}$$

which for integer $m \geq 0$ quickly become stationary and (13) is easily evaluated ($N \approx \frac{1}{2}(\frac{2}{m+1})^{\frac{1}{m+1}}$). The case $m = 0$ is the Has'minskiĭ process [18], which as $t \to \infty$ becomes an exponential distribution ($p(x, \infty) \sim exp(-2|x|)$); and $m = 1$ is the Ornstein-Uhlenbeck or classic Langevin process which is asymptotically Gaussian. Computing the left-hand sides of inequalities (17) and (19), it was easy to study the stability regions. Our monomial drift examples have symmetric distributions, so only $x_0 > 0$ is needed for illustration. Figures 1,2 plot the ratio

$$q = \frac{<|x_h|^2>}{|x_0|^2},$$

for various step sizes ($h = 1/10, 1/100$) and values $m = 2, 3, 4$. To lowest order in $h$ and small $|x_0|$, this ratio is $q \sim (|x_0|^2 + h)/|x_0|^2$. Note that $<|x_h|^2> - |x_0|^2 = (q-1) \cdot |x_0|^2$, so $q > 1$ for large $|x_0|$ indicates a diffusive or growing process, hence is unstable. Respectively, Figure 1 shows this function for the explicit algorithm (15) and Figure 2 that of (16). We note that clearly the implicit method keeps this function less than unity, and thus remains stable for all

$$|x_0| < \left(\frac{2}{(m-1)h}\right)^{\frac{1}{m-1}}.$$

This is the size limit at which the $O(h)$ diffusion term in (17), $|1 + \frac{h}{2}b'_0|^2 h$, vanishes. The explicit method (15) gives $<|x_h|^2>$ greater than $|x_0|^2$ when $|x_0|$ is large enough, and therefore becomes unstable.

Finally, in the quadratic ($m = 2$) and cubic ($m = 3$) cases, the implicit formula (16) was solved exactly for $x_h$. Expanding the quadratic and cubic solutions in a Taylor series (in $\xi$) permitted an independent comparison with the approximate form (19). Namely, the expectation $<|x_h|^2>$ was computed to the desired order in $<\xi^{2 \cdot k}> = O(h^k)$ needed to achieve reasonable accuracy (e.g. plotting accuracy). Such comparisons for the $m = 2, 3$ cases are also shown in Figure 2 (labeled X2.1 for $m = 2$ and $h = 0.1$, X2.01 for $m = 2$ and $h = 0.01$, etc.). These X2.1, X2.01 ($m = 2$), and X3.1, X3.01 ($m = 3$) exact solutions of the discretized equation (16) are to be compared with the 2.1, 2.01 and 3.1, 3.01 curves, respectively, which were computed using the approximate formula (19). We note that whenever $|x_0|$ is not too large, the approximation (19) gives quite good results, and is qualitatively correct for larger process values. *A fortiori*, the exact solutions shown by curves X2.1, X2.01, X3.1, and X3.01 actually show better stability than the approximation (19) calculated using (18) shown by curves 2.1, 2.01, 3.1, and 3.01 in Figure 2. For the monomial drifts, thus polynomial drifts, estimate (19) thus gratefully appears pessimistic.

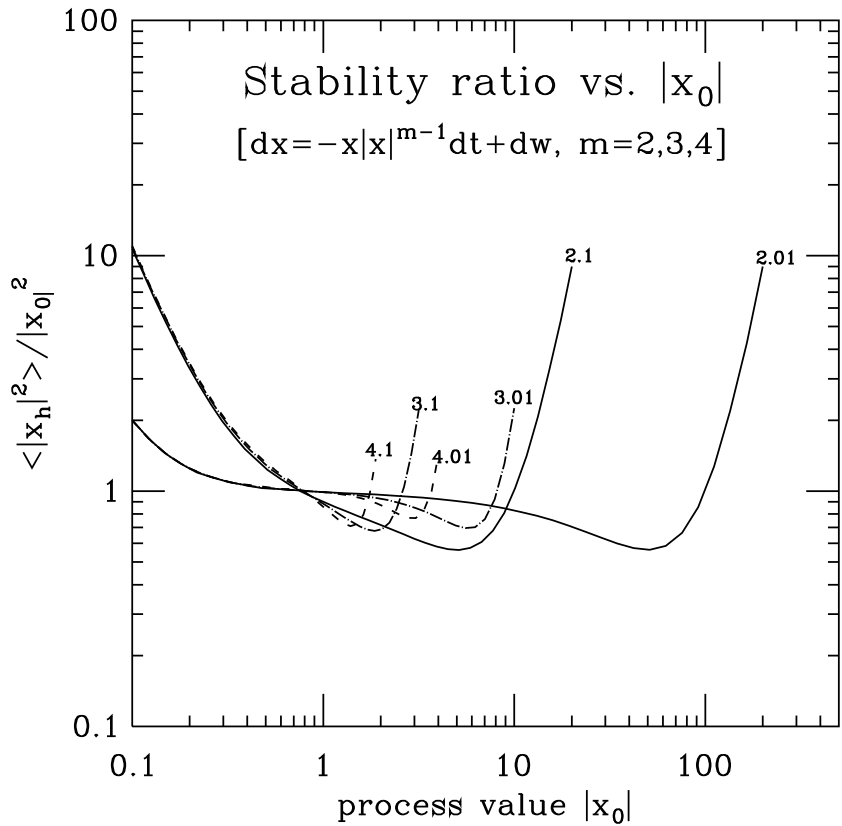

Figure 1: Stability of explicit trapezoidal rule (15) for problem (22). Curves labeled $m.d$ mean $m$ of (22) with stepsize $h = 0.d = 0.1$, or $0.01$.

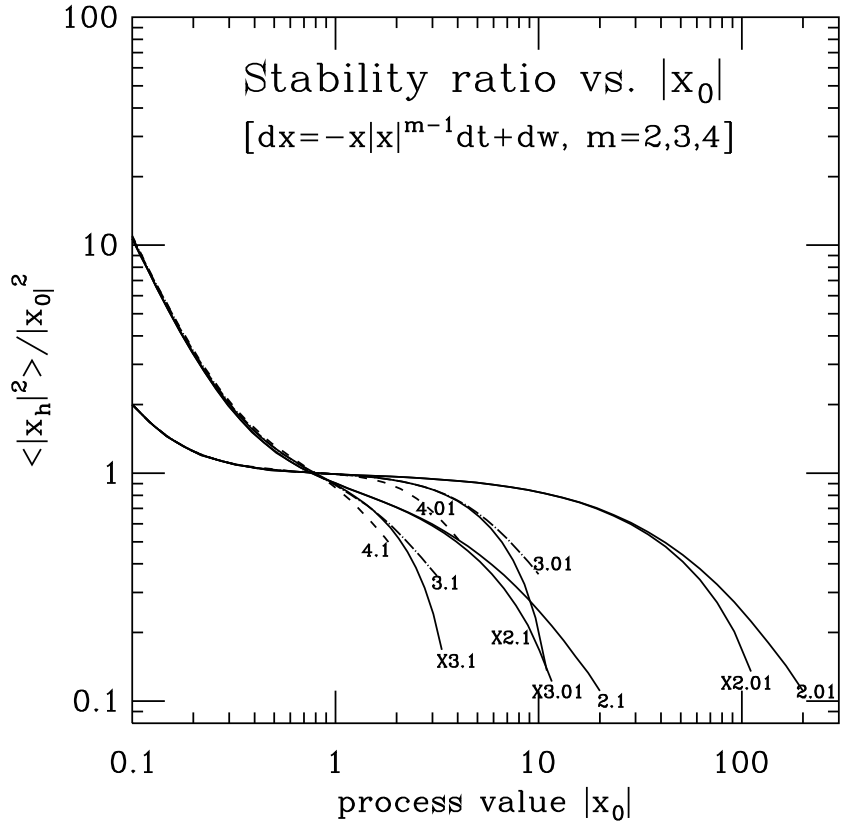

Figure 2: Stability of implicit trapezoidal rule (15) for problem (22). Curves labeled $m.d$ mean $m$ of (22) with stepsize $h = 0.d = 0.1$ or $0.01$. Labels $Xm.d$ refer to exact solutions of the implicit difference equation (16) and are to be compared to the approximate analysis (19).

# 5 Conclusions

We have shown a general splitting for a 2nd order weak accurate simulation method for Itô stochastic differential equations. The splitting permits choices for implicit dependences in the discretized time stepping which can improve stability. These choices can be made according to ease of solution (linear semi-implicit methods), or to improve stability when non-linear drift terms cause difficulties. Such methods have been shown useful in polymer physics [3]. Additionally, an approximate analysis of stability computed using the semi-implicit approximation (18) seems to yield quantitatively reliable predictions for additive noise problems when process sizes aren't too large, and seems qualitatively reliable in any case. This analysis shows that improvement in stability can be expected at least for polynomial non-linearities by using implicit trapezoidal rule. Finally, it has not escaped our notice that (18) is easily generalizable to the multiplicative noise case ($\sigma(x) \neq 1$), to yield a 2nd order weakly accurate linearly stabilized algorithm with no implicit equations to solve. The drawback of such a procedure being principally more functional evaluations (i.e. $b', b''$, if available), but will still be doubtlessly easier than solving implicit equations in the general case.

# 6 References

[1] W. P. Petersen, Journal of Computational Physics, vol. 113, no. 1, pp. 75-81, July 1994.

[2] W. P. Petersen, *Stability and Accuracy of Simulations for Stochastic Differential Equations*, IPS Research Report No. 90-02, Jan. 1990 (unpublished).

[3] H. C. Öttinger, *Stochastic Processes in Polymeric Fluids*, Springer-Verlag Berlin-Heidelberg, 1996.

[4] H. Schurz, *Numerical Regularization for SDEs: Construction of Non-negative Solutions*, Weirstrass Institute für Angewandte Analysis und Stochastik, pre-print no. 160, 1995. To be published in the Journal of Dynamical Systems and Applications.

[5] Georgio Parisi and Yong-Shi Wu, Sci. Sin. 24, p483 (1981).

[6] H. Gausterer and J.R. Klauder, Phys. Rev. D33, 3678, (1986).

[7] Kiosi Itô, "Differential Equations Determining Markov Processes," *Zenkoku Shijo Sugaku Danwakai*, vol. 244, no. 1077, pp. 1352-1400 (1942).

[8] P.E. Kloeden and E. Platen, *Numerical Solution of Stochastic Differential equations*, Springer Verlag, Berlin, 1992.

[9] G.N. Milstein, *Numerical Solution of Stochastic Differential equations*, revised and translated version of 1988 Russian work (published by Ural State University

Press), Kluwer Academic Publishers 1995.

[10] Diego Brico Hernandez and Renato Spigler, *Convergence and Stability of Implicit Runge-Kutta Methods for Systems with Multiplicative Noise*, BIT, vol. 33, pp. 654-669 (1993).

[11] Denis Talay, *Discrétisation d'une Équation Différentielle Stochastique et Calcul Approché d'Espérances de Fonctionelles de la Solution*, Mathematical Modeling and Numerical Analysis, vol. 20, no. 1, pp. 141-179, AFCET Gauthier-Villars, 1986.

[12] Ikeda and Watanabe, *Stochastic Differential Equations and Diffusion Processes*, Elsevier-North Holland Publ., Kodansha, 1981.

[13] R. Bulirsch and J. Stoer, *Introduction to Numerical Analysis*, Springer-Verlag, New York, 1980.

[14] C. W. Gear, *Numerical Initial Value Problems in Ordinary Differential Equations*, Prentice-Hall, Inc., Englewood Cliffs, NJ, 1971.

[15] W. Wagner and E. Platen, *Approximation of Itô Integral Equations*, Preprint ZIMM, Akad. Wissenschaft. DDR, Berlin, 1978.

[16] J. C. Butcher, Math. Comp., No. 18, pp. 50-64, 1964; J. C. Butcher, Math. Comp., No. 18, pp. 233-244, 1964; and J. C. Butcher, *The Numerical Analysis of Ordinary Differential Equations*, Wiley-Interscience Publ., 1987.

[17] J. L. Doob, *Stochastic Processes*, John Wiley & Sons, Inc., 1953.

[18] R. Z. Has'minskiĭ, *Stochastic Stability of Differential Equations*, Sijthoff and Noordhoff, Alphen aan den Rijn, The Netherlands, 1980